# Smart Cities Software from the developer's point of view

## Dmitry Namiot[1], Manfred Schneps-Schneppe[2]

[1]Faculty of Computational Mathematics and Cybernetics, Lomonosov Moscow State University, Moscow, Russia
dnamiot@gmail.com
[2]Ventspils International Radio Astronomy Centre Ventspils University College, Ventspils, Latvia
manfreds.sneps@gmail.com

**Abstract:** *The paper discusses the current state and development proposals for Smart Cities and Future Internet projects. Definitions of a Smart City can vary but usually tend to suggest the use of innovative Info-Communication technologies such as the Internet of Things and Web 2.0 to deliver more effective and efficient public services that improve living and working conditions and create more sustainable urban environments. Our goal is to analyze the current proposals from the developer's point of view, highlight the really new elements, the positions borrowed from the existing tools as well as propose some new extensions. We would like to discuss the possible extensions for the existing proposals and describe add-ons that, by our opinion, let keep the future research inline with the modern approaches in the web development domain.*

### Introduction

The concept of "smart cities" has attracted considerable attention in the context of urban development policies. The Internet and broadband network technologies as enablers of e-services become more and more important for urban development while cities are increasingly assuming a critical role as drivers of innovation in areas such as health, inclusion, environment and business (Schaffers et al., 2011). There is a limitless amount of available online resource and articles, devoted to smart cities and related IT tools. Our goal in this article is to analyze them from the software development point of view.

There are actually more than one defintion for the term „smart city". For example, IBM's Smart Cities digitize and connect infrastructures (IOT) to infuse them with new intelligence (IBM, 2012). As per Forrester, Smart City is the combined use of software systems, server infrastructure, network infrastructure, and client devices to better connect seven critical city infrastructure components and services: city administration, education, healthcare, public safety, real estate, transportation, and utilities (Smart City, 2012).

The rest of the paper is organized as follows. We consider two projects: EPIC project, and FI-Ware project. Then we analyze the Open API for M2M, submitted to ETSI, and offer a new web tool, Web Intents for M2M, as an enhancement of M2M middleware.

### EPIC project

Altogether, seven pilot projects were selected for the call for funding under the CIP ICT Policy Support Program on "Open Innovation for Future Internet-enabled services in smart cities". This list includes the following projects: EPIC, Smart IP, Peripheria, People, Open Cities, Life 2.0 and Smart-Islands.

The European Platform for Intelligent Cities (EPIC) is a European Commission-funded project that aims to wed state-of-the-art cloud computing technologies with fully researched and tested e-Government service applications to create the first truly scalable and flexible pan-European platform for innovative, user-driven public service delivery.

The EPIC platform will combine the industrial strength of IBM's 'Smart City' vision and cloud computing infrastructure with the knowledge and expertise of the Living Lab approach (which expressly engages citizens in service design) to ensure the development of a European 'innovation ecosystem' to deliver sustainable, user-driven web-based services for citizens and businesses.

The EPIC project will help to significantly accelerate the uptake of new citizen-generated services across Europe by combining the world-leading business expertise of Deloitte Consulting with the practical, first-hand knowledge of the European Network of Living Labs to guide cities through the routes, decisions and steps they need to undertake to improve service delivery and achieve the benefits of 'smart' working. Figure 1 presents the whole project.

The key points, from the development point of view, are: test and development cloud and its APIs, service catalogue and its API's (at least discovery and publishing), and IoT middleware.

At this moment, we can already highlight one significant moment: the word "middleware" actually describes the very wide area of software. And it is not understandable, for example, why cloud and directory of services are separated from the middleware? All three components could be presented actually as middleware.

In the terms of software, EPIC approach is based on the well known Web services approach in the traditional (canonical) form: WSDL, SOAP, etc. At this stage, EPIC does not introduce really new elements for the developers community.

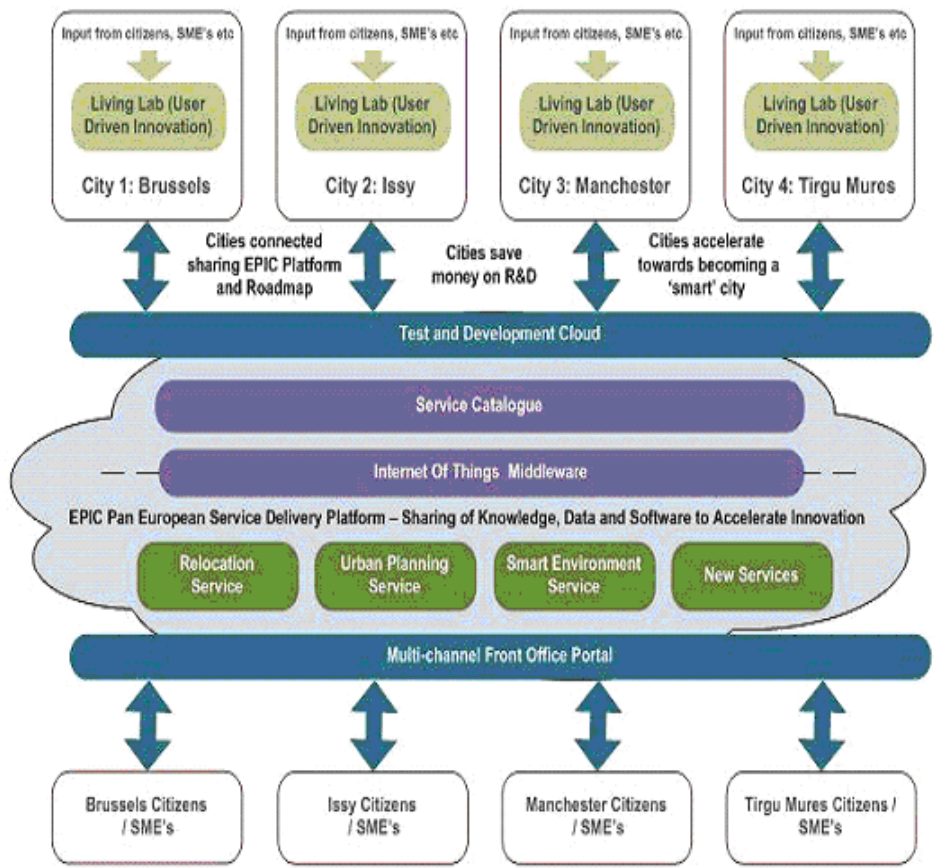

Fig. 1. **EPIC project roadmap** (EPIC, 2012).

## FI-Ware project

The more interesting from the developer's point of view is FI-Ware project. FI-WARE is being developed as part of the Future Internet Public Private Partnership program launched by the European Commission in collaboration with the ICT Industry.

The Reference Architecture of the FI-WARE platform is structured along a number of technical chapters, namely: Cloud Hosting, Data/Context Management, Internet of Things (IoT) Services Enablement, Applications/Services Ecosystem and Delivery Framework, Security and Interface to Networks and Devices. As per official document, FI-WARE will enable smarter, more customized/personalized and context-aware applications and services by the means of a set of assets able to gather, exchange, process and analyze massive data in a fast and efficient way (Figure 2).

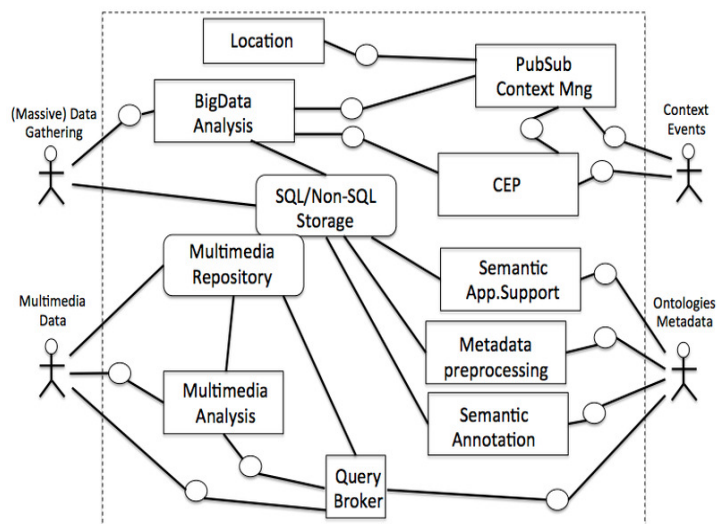

Fig. 2. **FI-WARE project** (FI-Ware, 2012a).

Data in FI-WARE refers to information that is produced, generated, collected or observed that may be relevant for processing, carrying out further analysis and knowledge extraction. A basic concept in FI-WARE is that data elements are not bound to a specific format representation. Actually the whole data model in FI-WARE has been described with the concept of NoSql systems in mind (Pokorny, 2011). Data items could be named and presents themselves by just named collection of triples: <name, type, value>.

What is important, that optionally, data elements could have meta-data (descriptions) associated with them. Meta-data elements could be described via collections of triples <name, type, value> too. The data-model described in FI-WARE is actually could be perfectly supported by distributed key-value systems (Lakshman, Malik, 2010).

Context in FI-WARE is represented through context elements. A context element extends the concept of data element by associating an *EntityId* and *EntityType* to it, uniquely identifying the entity (which in turn may map to a group of entities) in the FI-WARE system to which the context element information refers. In addition, there may be some attributes as well as meta-data associated to attributes that we may define as mandatory for context elements as compared to data elements (FI-Ware, 2012b).

It is very important that FI-WARE actually uses the same model for data and meta-data. It means that from the developer's point of view it should be possible to use the same model for persistence and search for both data and meta-data.

An event in FI-WARE is an act of creation a new element. It could be either data event (create data elements), or context event (creates context element). As an example, a sensor device measures some value and periodically creates and sends a new context element. The creation and sending of the context element is an event.

Because each event has got either data or context elements linked to, the whole system can see events via linked data. It makes the whole system much more uniform (homogeneous) comparing with M2M approach described below.

## Discussion

At this moment we have a wide choice for real-time analytical systems based on key-value stores. For example, we can mention Google Percolator (Peng, Dabek, 2010) or Twitter Storm (Marz, 2011). It is exactly the approach needed for processing data in FI-WARE model. For event publishing, FI-WARE roadmap suggests ContextML (Knappmeyer et al., 2010) and SPARQL (Sparql, 2012).

*ContextML* is light-weight XML based context representation schema in which context information is categorized into scopes and related to different types of entities (e.g. user, device). The schema is also applied for encoding management messages in order to allow for a flexible framework supporting gradual plug & play extendibility and mobility. *ContextML* is tailored to be used for REST-based communication between the framework components.

*SPARQL.* RDF is a directed, labeled graph data format for representing information in the Web. And SPARQL specification defines the syntax and semantics of the query language for RDF. SPARQL can be used to express queries across diverse data sources, whether the data is stored natively as RDF or viewed as RDF via middleware. SPARQL contains capabilities for querying required and optional graph patterns along with their conjunctions and disjunctions. SPARQL also supports extensible value testing and constraining queries by source RDF graph. The results of SPARQL queries can be results sets or RDF graphs.

*O&M example.* The obvious candidates here are standards activities from The Open Geospatial Consortium (OGC) that focus on sensors and sensor networks comprise (Botts, 2008). At the first hand it is Observations & Measurements Schema (O&M) as well as Sensor Model Language (SensorML), Transducer Model Language (TransducerML or TML), Sensor Observations Service (SOS), Sensor Planning Service (SPS) and Sensor Alert Service (SAS).

For example O&M supports data sampling as this:

<gml:description>

  Observation test instance: fruit mass

</gml:description>

 <gml:name>Observation test 1</gml:name>

  <om:phenomenonTime>

 <gml:TimeInstant   gml:id="ot1t">

<gml:timePosition>

2005-01-11T16:22:25.00

</gml:timePosition>

   </gml:TimeInstant>

 </om:phenomenonTime>

  <om:parameter>

<om:NamedValue>

<om:name xlink:href="http://sweet.jpl.nasa.gov/ontology/property.owl#Temperature"/>

```
<om:value xsi:type="gml:MeasureType" uom="Cel">
```

22.3

```
</om:value>
```

```
</om:NamedValue>
```

```
</om:parameter>
```

*XML vs. JSON.* But keeping in mind the modern trend in web development – shall we keep that as XML, or it is a time to replace it with an appropriate JSON?

FI-WARE proposes also an interesting approach for Applications/Services Ecosystem and Delivery Framework. It is based on the heavy usage on USDL (Barros, 2012). Universal Service-Semantics Description Language (USDL) can be used by service developers to specify formal semantics of web-services. Thus, if WSDL can be regarded as a language for formally specifying the syntax of web-services, USDL can be regarded as a language for formally specifying their semantics. USDL is as formal service documentation that will allow sophisticated conceptual modeling and searching of available web-services, automated composition, and other forms of automated service integration. For example, the WSDL syntax and USDL semantics of web services can be published in a directory which applications can access to automatically discover services (Park, 2011).

## Open API for M2M

Why we think that it is a must to talk about M2M API here? Actually, in the many aspects, IoT and M2M applications are performing absolutely the same tasks. For example, both systems could be described as data collectors for the future processing. This section describes an Open API for M2M, submitted to ETSI. It is probably the most valuable achievement at this moment. Roadmap for FI-WARE highlights the plans for M2M General Enabler (Gateway). In the same time M2M Open API can benefits from more elaborated FI-WARE persistence and cloud offerings.

The Open API for M2M applications are developed jointly in EURESCOM study P1957 and the EU FP7 SENSEI project. The Open API has been submitted as a contribution to ETSI TC M2M for standardization (ETSI, 2011).

Actually, in this Open API, we can see the big influence of Parlay specification. Parlay Group leads the standard, so called Parlay/OSA API, to open up the networks by defining, establishing, and supporting a common industry-standard APIs. Parlay Group also specifies the Parlay Web services API, also known as Parlay X API, which is much simpler than Parlay/OSA API to enable IT developers to use it without network expertise (Yim, 2006).

The goals are obvious, and they are probably the same as for any unified API. One of the main challenges in order to support easy development of M2M services and applications will be to make M2M network protocols "transparent" to applications. Providing standard interfaces to service and application providers in a network independent way will allow service portability (Gronbek, 2008).

At the same time, an application could provide services via different M2M networks using different technologies as long as the same API is supported and used. By this way, an API shields applications from the underlying technologies and reduces efforts involved in service development. Services may be replicated and ported between different execution environments and hardware platforms (Gronbek, Ostendorf, 2010).

This approach also lets services and technology platforms to evolve independently. A standard open M2M API with network support will ensure service interoperability and allow ubiquitous end-to-end service provisioning.

The Open API provides service capabilities that are to be shared by different applications. Service Capabilities may be M2M specific or generic, i.e., providing support to more than one M2M application.

Key points for Open API are:
- it supports interoperability across heterogeneous transports
- ETSI describes high-level flow and does not dictate implementation technology
- it is message-based solution
- it combines P2P with client-server model
- and it supports routing via intermediaries

At this moment, all point are probably not discussable except the message-based decision. Nowadays, publish-subscribe method is definitely not among the favorites approaches in the web development, especially for heavy-loading projects.

Let us name the main Open API categories (Table 1) and make some remarks (Sneps-Sneppe, Namiot, 2012a).

Table 1

## Open API categories and remarks

| Open API categories | API contents | Comments |
|---|---|---|
| *Grouping* | A group here is defined as a common set of attributes (data elements) shared between member elements. On practice it is about the definition of addressable and exchangeable data sets. | Just note, as it is important for our future suggestions, there are no persistence mechanisms for groups. |
| *Transactions* | Service capability features and their service primitives optionally include a transaction ID in order to allow relevant service capabilities to be part of a transaction. Just for the deploying transactions and presenting some sequences of operations as atomic. | In the terms of transactions management Open API presents the classical 2-phase commit model. By the way, we should note here that this model practically does not work in the large-scale web applications. We think it is very important because without scalability we |

| | | cannot think about "billions of connected devices". |
|---|---|---|
| *Application Interaction* | The application interaction part is added in order to support development of simple M2M applications with only minor application specific data definitions: readings, observations and commands. | Application interactions build on the generic messaging and transaction functionality and offer capabilities considered sufficient for most simple application domains. |
| *Messaging* | The Message service capability feature offers message delivery with no message duplication. Messages may be unconfirmed, confirmed or transaction controlled. | The message modes supported are single Object messaging, Object group messaging, and any object messaging; (it can also be Selective object messaging). Think about this as Message Broker. |
| *Event notification and presence* | The notification service capability feature is more generic than handling only presence. It could give notifications on an object entering or leaving a specific group, reaching a certain location area, sensor readings outside a predefined band, an alarm, etc. | It is a generic form. So, for example, geo fencing should fall into this category too. The subscriber subscribes for events happening at the Target at a Registrar. The Registrar and the Target might be the same object. This configuration offers a publish/subscribe mechanism with no central point of failure. |
| *Compensation* | Fair and flexible compensation schemes between cooperating and competing parties are required to correlate resource consumption and cost, e.g. in order to avoid anomalous resource consumption and blocking of incentives for investments. The defined capability feature for micro-payment additionally allows charging for consumed network resources. | It is very similar, by the way, to Parlay's offering for Charging API. |
| *Sessions* | In the context of OpenAPI a session shall be understood to represent the state of active communication between Connected Objects. | OpenAPI is REST based, so, the endpoints should be presented as some URI's capable to accept (in this implementation) the basic commands GET, POST, PUT, DELETE (See an example below). |

*A session example:* requests execution of some function.

URI: http://{nodeId}/a/do

Method: POST

Request

<?xml version="1.0" encoding="UTF-8" standalone="yes"?>

<appint-do-request xmlns="http://eurescom.eu/p1957/openm2m">

<requestor>9378f697-773e-4c8b-8c89-27d45ecc70c7</requestor>

<commands>

<command>command1</command>

<command>command2</command>

</commands>

<responders>9870f7b6-bc47-47df-b670-2227ac5aaa2d</responders>

<transaction-id>AEDF7D2C67BB4C7DB7615856868057C3</transaction-id>

</appint-do-request>

Response

<?xml version="1.0" encoding="UTF-8" standalone="yes"?>

<appint-do-response xmlns="http://eurescom.eu/p1957/openm2m">

<requestor>9378f697-773e-4c8b-8c89-27d45ecc70c7</requestor>

<timestamp>2010-04-30T14:12:34.796+02:00</timestamp>

<responders>9870f7b6-bc47-47df-b670-2227ac5aaa2d</responders>

<result>200</result>

</appint-do-response>

Note that because we are talking about server-side solution, there is no problem with so called sandbox restrictions. But it means of course, that such kind of request could not be provided right from the client side as many modern web applications do. In the same time FI-WARE authors, at least according to roadmap, pays attention to JSON and other client-side technologies. The reasons are obvious – the modern web-based technologies can save the development time. The next section below is also devoted to client-side technologies.

*Data persistence.* Another area, where M2M API in the current form is weak by our opinion, belongs to data persistence. We should keep in mind that we are talking about the particular domain – M2M. In the most cases, all business applications here will deal with some metering data. As soon as we admit, that we are dealing with the measurements in the various forms, we should make, as seems to us a natural conclusion – we need to save the data somewhere. It is the core business for M2M – save data for the future processing.

So, the question is very easy – can we talk about M2M applications without talking about data persistence? Again, the key question is M2M. It is not some abstract web API. We talk about the well-defined domain here.

As seems to us, even right now, before the putting some unified API in place, the term M2M almost always coexists with the term "cloud". And as we can see, almost always has been accompanied by the terms like automatic database logging, backup capabilities, etc.

So, maybe this question is more for the discussions or it even could be provocative in the some forms, but it is: why there is no reference API for persistence layer in the unified M2M API? It is possible in general to create data gathering API without even mentioning data persistence?

## Web intents vs. Open API from ETSI

This section is devoted to the relatively new approach in the client-side web development – Web Intents. The first time Web Intents usage for M2M applications was proposed by Dmitry Namiot (Sneps-Sneppe, Namiot, 2012b).

Let us start from the basic. Users use many different services on the web to handle their day to day tasks, developers use different services for various tasks. In other words, our environment consists of connected applications. And of course, all they expect their applications to be connected and to work together seamlessly.

It is almost impossible for developers to anticipate every new service and to integrate with every existing external service that their users prefer, and thus, they must choose to integrate with a few select APIs at great expense to the developer.

As per telecom experience, we can mention here the various attempts for unified API that started, probably, with Parlay. Despite a lot of efforts, Parlay API's actually increase the time for development. It is, by our opinion, the main reason for the Parlay's failure.

Web Intents solves this problem. Web Intents is a framework for client-side service discovery and inter-application communication. Services register their intention to be able to handle an action on the user's behalf. Applications request to start an action of a certain verb (for example: share, edit, view, pick, etc.) and the system will find the appropriate services for the user to use based on the user's preference. It is the basic (Namiot, 2013).

Going to M2M applications it means that our potential devices will be able to present more integrated for the measurement visualization for example. The final goal of any M2M based application is to get (collect) measurements and perform some calculations (make some decisions) on the collected dataset. We can go either via low level API's or use (at least for majority of use cases) some integrated solutions. The advantages are obvious. We can seriously decrease the time for development.

*Web Intents example.* Web intents put the user in control of service integrations and make the developers life simple. Here is the modified example for web intents integration for the hypothetical web intents example:

1. Register some intent upon loading our HTML document

document.addEventListener("DOMContentLoaded", function() {

    var regBtn = document.getElementById("register");

    regBtn.addEventListener("click", function() {

    window.navigator.register("http://webintents.org/m2m", undefined);

    }, false);

2. Start intent's activity

    var startButton = document.getElementById("startActivity");

    startButton.addEventListener("click", function() {

     var intent = new Intent();

     intent.action = "http://webintents.org/m2m";

      window.navigator.startActivity(intent);

    }, false);

3. Get measurements (note – in JSON rather than XML) and display them in our application

    window.navigator.onActivity = function(data) {

```
    var output = document.getElementById("output");

    output.textContent = JSON.stringify(data);

  };

}, false);
```

*Discussion.* Obviously, that it is much shorter than the long sequence of individual calls as per M2M Open API. The key point here is *onActivity* callback that returns JSON (not XML!) formatted data. As per suggested M2M API we should perform several individual requests, parse XML responses for the each of them and only after that make some visualization. Additionally, web intents based approach is asynchronous by its nature, so, we do not need to organize asynchronous calls by our own.

Also, Web Intents approach lets us bypass sandbox restrictions. In other words, developers can raise requests right from the end-user devices, rather than always call the server. The server-side only solution becomes bottleneck very fast. And vice-versa, client side based requests let developers deploy new services very quickly. Why do not use the powerful browsers in the modern smart-phones? At the end of the day, Parlay specifications were born in the time of WAP and very rudimentary mobile browsers. Why do we ignore HTML5 browsers and JavaScript support in the modern phones?

**Future work**

Considering M2M communications as a central point of Future Internet, European commission creates standardization mandate M/441. The Standardization mandate M/441, issued on 12th March 2009 by the European mandate to CEN, CENELEC and ETSI, in the field of measuring instruments for the development of an open architecture for utility meters involving communication protocols enabling interoperability, is a major development in shaping the future European standards for smart metering and Advanced Metering Infrastructures. The general objective of the mandate is to ensure European standards that will enable interoperability of utility meters (water, gas, electricity, heat), which can then improve the means by which customers' awareness of actual consumption can be raised in order to allow timely adaptation to their demands.

ETSI's 3rd workshop on Machine to Machine (M2M) communications, held in Mandelieu, France on October 23-25, 2012, gathered leading experts from all over the world to hear how ETSI M2M technology standards are being deployed. "With 270 registered delegates from four continents, 25 speakers, thirteen live demonstrations of M2M-based applications and two days of intense discussion, this year's event was again a success," – the official workshops' site assure us (M2MWORKSHOP, 2012). In reality, the state-of-the-art with M2M standards is far from hopeful.

The demos covered once again a respectable cross section of the application domains such as: Smart Metering, Home automation, Energy Efficiency, Smart building, Smart City, Smart Parking, Exercise, Gaming and Home Energy, Management Systems linked with Social Networking Service and others. But the existing standards CoAP, 6lowpan, ETSI M2M, OMA DM, BBF TR069, OSGi, HGI, etc. and the ZigBee, KNX, etc. are far from convergence.

The new international M2M Partnership Project "oneM2M" is now started. The list of funding partners include ETSI (Europe), ATIS and TIA (US), CCSA (China), TTA (Korea), ARIB and TTC (Japan). And the leading role of ETSI goes more sophisticated.

**Conclusions**

This paper discusses the current state and development proposals for Smart Cities and Future Internet projects. Our goal was to analyze the current proposals from the developer's point of view. Software development companies and universities should start investigate tools used and proposed in the Future Internet roadmap right now. Also we discussed the possible extensions for the existing proposals and describe the add-ons that, by our opinion, let keep the future research inline with the modern approaches in the web development domain. Article proposes some new additions – web intents as add-on for the more traditional REST approach. The main goal for our suggestions is the simplifying of the development phases for new applications by support asynchronous calls and JSON versus XML data transfer.

**Acknowledgments**

The paper is financed from ERDF's project SATTEH (No. 2010/0189/2DP/2.1.1.2.0/10/APIA/VIAA/019) being implemented in Engineering Research Institute «Ventspils International Radio Astronomy Centre» of Ventspils University College (VIRAC).